\documentclass[epj]{svjour}

\usepackage{amsmath}
\usepackage{amssymb}
\usepackage{graphicx}
\usepackage{dcolumn}
\usepackage{bm}
\usepackage{multirow}
\usepackage{booktabs}

\usepackage{subfigure}

\graphicspath{{images/}{plots/}}

\newcommand{\avg}[1]{\ensuremath{\left< #1 \right>}}

\newcommand{\abs}[1]{\ensuremath{\left\vert#1\right\vert}}

\newcommand{\de}{\,\mathrm{d}}

\renewcommand{\vec}[1]{\bm{#1}}

\DeclareMathOperator{\ee}{e}

\begin{document}
\title{Large-deviation properties of the largest biconnected component for random graphs}

\author{Hendrik Schawe%
    \thanks{\emph{Present address:} hendrik.schawe@uni-oldenburg.de}%
    \and Alexander K. Hartmann%
    \thanks{\emph{Present address:} a.hartmann@uni-oldenburg.de}
}
\institute{Institut f\"ur Physik, Universit\"at Oldenburg, 26111 Oldenburg, Germany}
\authorrunning{H. Schawe \and A. K. Hartmann}
\date{\today}

\abstract{
    We study the size of the largest biconnected components in sparse
    Erd\H{o}s-R\'enyi graphs with finite connectivity and Barab\'asi-Albert
    graphs with non-integer mean degree. Using a statistical-mechanics
    inspired Monte Carlo approach we obtain numerically
    the distributions for different sets of parameters over almost their whole
    support, especially down to the rare-event tails with probabilities far
    less than $10^{-100}$.
    This enables us to observe a qualitative difference in the behavior of
    the size of the largest biconnected component and the largest $2$-core
    in the region of very small components, which is unreachable using
    simple sampling methods. Also, we observe a
    convergence to a rate function even for small sizes, which is a hint
    that the large deviation principle holds for these distributions.
}

\maketitle

    \section{Introduction}
        The robustness of networks \cite{newman2003review,dorogovtsev_book2006,newman_book2006,newman_book2010,barrat_book2012}
        attracted much interest in recent time,
        from practical applications for, e.g., power grids~\cite{Sachtjen2000Disturbances,Rohden2012self,dewenter2015large},
        the internet~\cite{Cohen2000Resilience,lee2006maximum},
        to examinations of genomes~\cite{ghim2005lethality,kim2015biconnectivity}.
        As typical in network science, one does not only study the
        properties of existing networks. To model the properties of real networks,
        different ensembles of random graphs were
        devised, e.g., Erd\H{o}s-R\'enyi random graphs~\cite{erdoes1960},
        small world graphs~\cite{watts1998}, or scale-free graphs~\cite{barabasi1999emergence}.
        Also for such ensembles the robustness has been studied by analytical and numerical
        means~\cite{albert2000error,Callaway2000network,Newman2008Bicomponents,Norrenbrock2016fragmentation}.
        One often used approach to determine the robustness of networks are
        \emph{fragmentation} studies, where single nodes are removed from the
        network. These nodes are selected according to specific rules (``attack'')
        or randomly (``failure''). The functionality,
        e.g., whether it is still connected, is tested afterwards. A property
        necessary for robustness is thus that the graph stays connected when
        removing an arbitrary node. This exact concept is characterized by the
        \emph{biconnected component}, which are the connected components which
        stay connected after an arbitrary node is removed. The existence of a
        large biconnected component is thus a simple and fundamental property of a graph robust
        to fragmentation. Another, though related, often studied form of stability
        looks at the flow through or the transport capability~\cite{lee2006maximum}
        of a graph. Also here a large biconnected component is a good indicator
        for stability. Intuitively, in a biconnected component there is never
        a single bottleneck but always a backup path to reach any node. This
        ensures the function of the network even in case that an arbitrary edge
        has too low throughput or an arbitrary node of the biconnected component
        is damaged.

        At the same time the biconnected component is a simple concept enabling
        to some extent its treatment by analytical means for some
        graph ensembles. For example, the mean size $\avg{S_2}$
        of the biconnected component for a graph with a given degree distribution
        is known~\cite{Newman2008Bicomponents}. Also the percolation transition
        of the biconnected component for scale free and Erd\H{o}s-R\'enyi graphs
        is known to coincide with the percolation transition of the single
        connected component and its finite size scaling behavior is
        known~\cite{Kim2013phase}. Nevertheless, a full description of any
        random variable is only obtained if its full probability distribution is known.
        To our knowledge, concerning the size of the biconnected component
        this has not been achieved so far for any graph ensemble, neither analytically
        nor numerically.

        For few network observables and some graphs ensembles results concerning
        the probability distributions have been already obtained so far.
        For the size of the connected component on Erd\H{o}s-R\'enyi random
        graphs analytical results~\cite{biskup2007large} for the rate function
        exist, i.e., the behavior of the full distribution for large
        graph sizes $N$. Numerically it was shown that this is already for
        relatively small $N$ a very good approximation~\cite{Hartmann2011large}.
        Corresponding numerical results for two-dimensional percolation
        have been obtained as well~\cite{Hartmann2011large}.
        Similarly there are numerical, but no analytical works, scrutinizing
        the size of the related $2$-core over most of its support
        again for Erd\H{o}s-R\'enyi random graphs~\cite{Hartmann2017Large}.

        Since similar results seem not to be available concerning the
        biconnected components, and given its importance for network robustness, this
        is an omission that we will start to cure with this study.
        Here, we numerically obtain the probability density function of
        the size of the largest biconnected component over a large part of its
        support, i.e., down to probabilities smaller than $10^{-100}$. This
        enables us also to directly observe large deviation properties, and
        shows strong hints that the large deviation principle
        holds \cite{denHollander2000,touchette2009} for this distribution.

        The remainder of this manuscript gives definitions of the graph
        ensembles and the properties of interest, as well as some known results,
        in section~\ref{sec:biconnected} and explains the sampling methods
        needed to explore the tails of the distributions in section~\ref{sec:sampling}.
        The results of our simulations and a discussion will follow in section~\ref{sec:results}.
        Section~\ref{sec:conclusion} summarizes the results.

    \section{Models and Methods\label{sec:mm}}

    \subsection{Biconnected Components of Random Graphs\label{sec:biconnected}}
        A \emph{graph} $G = (V, E)$ is a tuple of a set of nodes $V$ and edges
        $E \subset V^{(2)}$. A pair of nodes $i, j$ are called \emph{connected},
        if there exists a \emph{path} of edges $\{i, i_1\},\{i_1, i_2\}, .., \{i_{k-1}, i_k\}, \{i_k, j\}$
        between them. A \emph{cycle} is a closed path, i.e., the edge $\{i,j\}$ exists
        and $i$ and $j$ are connected in $G^\prime = (V, E \setminus \{i,j\})$.
        The \emph{connected components} are the maximal disjoint
        subgraphs, such that all nodes of each subgraph are connected.

\begin{figure}[bhtp]
            \centering
            \includegraphics[width=0.4\textwidth]{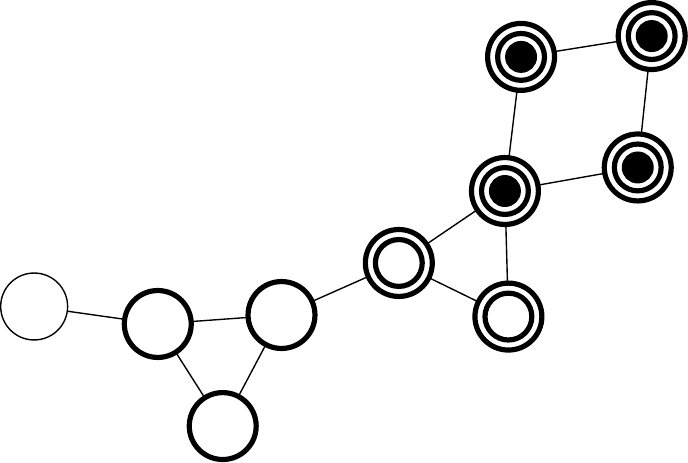}
            \caption{\label{fig:component_example}
                Every node is part of the connected
                component, nine nodes with bold borderline are
                part of the $2$-core, six nodes containing a circle
                are part of the largest bi-edge-connected component and all
                nodes containing a black dot are part of the largest
                biconnected component.
            }
        \end{figure}

        A \emph{biconnected component} (sometimes \emph{bicomponent}) of an
        undirected graph is a subgraph, such that every node can be reached by
        two paths, which are distinct in every node except the start- and end
        node.
        Thus, if any single node is removed
        from a biconnected component it will still be a connected component.
        Therefore clearly, each biconnected component is a connected component.
        We will also look shortly at \emph{bi-edge-connected components}, which are very
        similar, but the two paths may share nodes as long as they do not share
        any edge. Note that a biconnected component is always a bi-edge-connected
        component, but the reverse is not necessarily true. An example
        is shown in Fig.~\ref{fig:component_example}.
        In this study we will study only the largest biconnected component $G_\mathrm{bi}$
        of two types of random graphs. Note that, while every biconnected component
        is also a connected component, the largest biconnected component does
        not need to be a subgraph of the largest connected component
        $G_\mathrm{cc}$, it may be part of another, smaller, connected component.
        However, its size $S_2 = |G_\mathrm{bi}|$ is always smaller or equal
        than the size of the largest connected component $S = \abs{G_\mathrm{cc}}$.
        Similarly, the size $S_\text{2-core}$ of the largest connected
        component of the \emph{$2$-core},
        the subgraphs that remain after iterative removal of all nodes with degree
        less than $2$, is an upper bound on $S_2$, since the $2$-core of a graph
        consists of bicomponents possibly linked by single edges.
        In Fig.~\ref{fig:component_example} the largest components of each type
        are visualized for an example connected graph.
        In fact for the sizes of the largest
        of the above introduced subgraphs, the following relation holds.
        \begin{align}
            \label{eq:order}
            S \ge S_\text{2-core} \ge S_\text{2-edge} \ge S_2.
        \end{align}
        As we will see below, for the ensemble of Erd\H{o}s-R\'enyi random graphs
        in the percolating phase, the distributions of $S_\text{2-core}, S_\text{2-edge}$,
        and $S_2$ are actually very similar to each other. One has indeed to inspect the
        far tails of the distributions to see differences, which also justifies that we
        study the large-deviation properties here.
        For the ensemble of Barab\'asi-Albert graphs we study, the same is true.
        While the distributions of $S_\text{2-core}$ and $S_2$ look very similar
        in the main region, a qualitative difference is observable in the tail
        of small components. The difference is even more pronounced than for the
        ER case, since the general form of the distribution changes qualitatively
        to a convex shape for $P(S_\text{2-core})$.

        The classical way to find biconnected components of a
        graph~\cite{Hopcroft1973algorithm}
        is based on a depth first search and thus runs in linear time.
        For each connected component a depth first search is started at an
        arbitrary root node of that component.
        For each node the current \emph{depth} of the search, i.e., at which
        level in the tree traversed by the depth first search the node is located,
        and the \emph{lowpoint} are saved. The lowpoint is the
        minimum of the depth of the neighbors (in the graph)
        of all descendants of the node (in the tree).
        Iff the depth of a node is less or equal the lowpoint of one of its
        children (in the tree), this node separates two biconnected components and is called
        \emph{articulation point}. For the root node of the search there is an
        exception. It is an articulation point, iff it has more than one child.
        The articulation points separate
        biconnected components and are members of all biconnected components
        separated by them. A better illustrated explanation can be found in
        \cite{cormen2009introduction}.
        After finding all biconnected components, we measure the size of the largest.
        We did not implement an algorithm, instead we simply used
        the efficient implementation provided by the LEMON graph library~\cite{lemon}.

        The mean size of the biconnected component of graphs with a given degree
        distribution $p_k$ is known for large graphs~\cite{Newman2008Bicomponents,ghoshal2009structural}
        to be
        \begin{align}
            \label{eq:bi_analytic}
            \avg{S_2} = 1 - G_0(u) - (1-u)G_0^\prime(u),
        \end{align}
        where $G_0(z) = \sum_k p_k z^k$ is the probability generating function,
        $G_0^\prime$ its derivative and $u$ the probability to reach a node
        not part of the giant connected component when following an edge. $u$ is
        determined by the solution of
        \begin{align}
            \label{eq:u}
            u = \sum_{k=0}^\infty q_k u^k,
        \end{align}
        with the excess degree distribution $q_k = (k+1) p_{k+1} / \avg{k}$.
        Knowing the degree distribution of Erd\H{o}s-R\'enyi graphs $G(N,p)$ to be
        \begin{align}
            p_k = \binom{N-1}{k} p^{k} (1-p)^{N-1-k},
        \end{align}
        allows the numerical evaluation of Eq.~\eqref{eq:u}. We will compare
        these predictions to our simulational results to scrutinize the behavior
        for finite $N$.

        The ensemble of \emph{Erd\H{o}s-R\'enyi} (ER) graphs $G(N, p)$ consists of
        $N$ nodes and each of the $N(N-1)/2$ possible edges occurs with probability $p$. The connectivity
        $c = Np$ is the average number of incident edges per node, the average
        \emph{degree}. At $c_c=1$ this ensemble shows a \emph{percolation transition}.
        That is in the limit of large graph sizes $N$ the size of the largest
        connected component is of order $\mathcal{O}(N)$ above this threshold
        and of order $\mathcal{O}(1)$ below. Interestingly this point is also
        the percolation transition of the biconnected component~\cite{Kim2013phase}.

        To a lesser extend we also study \emph{Barab\'asi-Albert} (BA) graphs~\cite{barabasi1999emergence}.
        The ensemble
        of BA graphs is characterized by a tunable mean degree $\avg{k}$ and its degree
        distribution follows a powerlaw $p(k) \propto k^{-3}$. Realizations are
        constructed using a growth process. Starting from a fully connected
        subgraph of $m_0$ (here $m_0 = 3$) nodes, in every iteration one more
        node is added and connected to $m \le m_0$ existing nodes $j$ with a
        probability $p_j \propto k_j$ dependent on their degree $k_j$ until
        the size of the graph is $N$. The parameter $m = 2\avg{k}$ by construction.
        Since $m = 1$ will always result in a tree, which is not biconnected at all and
        $m \ge 2$ will always be a full biconnected component, we will allow
        fractional $1 < m < 2$ in the sense, that one edge is always added and
        a second with probability $m-1$.

    \subsection{Sampling\label{sec:sampling}}
        Since we are interested in the far tail behavior of the distribution of
        the size of the largest biconnected component, it is unfeasible to use
        naive simple sampling, i.e., uniformly generating configurations,
        measuring the observable and constructing a histogram. Instead we use
        a Markov chain Monte Carlo based importance sampling scheme to collect
        good statistics also in the far tails.
        This technique was already applied to obtain the distributions over a large
        range for the score of sequence alignments~\cite{Hartmann2002Sampling,Wolfsheimer2007local,Fieth2016score}, 
        to obtain statistics of the convex hulls of a wide range of types of random walks~\cite{Claussen2015Convex,Dewenter2016Convex,schawe2018avoiding,schawe2018large},
        to work distributions for non-equilibrium systems~\cite{Hartmann2014high}
        and especially to different properties of Erd\H{o}s-R\'enyi random graphs~\cite{Engel2004,Hartmann2011large,Hartmann2017Large,hartmann2018distribution}.

        The Markov chain in this case is a chain of random number vectors
        $\vec{\xi}_t$, $t=1,2,\ldots$.  Each entry of $\vec{\xi}_t$
        is drawn from a uniform $U(0,1)$ distribution. Each vector serves as an input
        for a function which generates a random graph. Since all randomness is included in
        $\vec{\xi}_t$, the generated graph $G_t=G(\vec{\xi}_t)$ depends deterministically on
        $\vec{\xi}_t$. In this way, the Markov chain $\{\vec{\xi}_t\}$ corresponds to a Markov
        chain $\{G_t\}$ of graphs. This approach, of separating the randomness from the
        actually generated objects, has the advantage that for the Markov chain
        we can generate graph
        realizations of arbitrary ensembles from scratch, without having to
        invent a valid Markov chain change move for each ensemble. However, for
        Erd\H{o}s-R\'enyi graphs, we use a specialized change move for performance
        reasons. One change move is to select a random node $i$, delete all
        incident edges and add every edge $\{i,j\}$ with $j \in V \setminus \{i\}$
        with probability $p$. For the Barab\'asi-Albert graphs such a simple change
        move is not trivial to construct. Therefore, for this type, we perform the typical growth
        process from scratch after changing one of the underlying random
        numbers in $\vec{\xi}_t$.

        \begin{figure}[bhtp]
            \centering
            \includegraphics[width=0.47\textwidth]{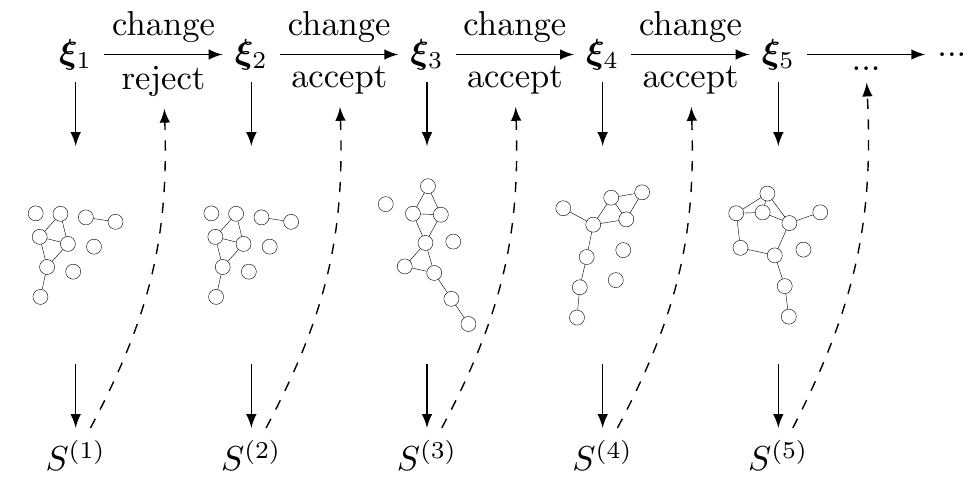}
            \caption{\label{fig:blackBox}
                Four steps of our importance sampling scheme at a small
                negative temperature, biasing towards a larger biconnected
                component.
            }
        \end{figure}

        The main idea to obtain
        good statistics over a large part of the support, especially for
        probabilities smaller than, say, $10^{-100}$, is to bias the generated
        samples towards those regions. Therefore, we will use classical
        Metropolis sampling to gather realizations of graphs $G$.
        The Markov chain underlying this method consists of either Graph
        realizations $G$ (ER case) or random number vectors $\vec\xi$ from
        which a graph realization can be constructed $G(\vec\xi)$ (BA case).
        We will describe the process for the latter more general case.
        We start our Markov chain with some random state $\vec\xi_1$ and
        at every iteration we propose a new state $\vec\xi^\prime$, i.e., replace a single entry
        of $\vec{\xi}_t$ with a new uniform random number and generate a new
        realization $G(\vec{\xi}^\prime)$ from these random numbers.
        We will accept this proposal as the new state $\vec{\xi}_{i+1}$, with
        the classical Metropolis acceptance probability
        $p_\mathrm{acc} = \min\{1,\ee^{-\Delta S/T}\}.$
        This process is sketched in Fig.~\ref{fig:blackBox}.
        Since we are interested in the size of the largest biconnected component
        $S$, we will treat this observable as the ``energy'' of the realization.
        Thus, $\Delta S$ is the difference in energy between the old and proposed
        state. Otherwise the proposal is rejected, i.e., $\vec{\xi}_{t+1} = \vec{\xi}_t$.
        Following this protocol, the Markov chain will equilibrate eventually and
        from thereon yield realizations $G(\vec\xi)$ which are
        Boltzmann distributed with respect to some ``artificial temperature'' $T$
        \begin{align}
            \label{eq:qt}
            Q_T(G) = \frac{1}{Z_T} \ee^{-S(G)/T} Q(G),
        \end{align}
        where $Q(G)$ is the natural distribution of the realizations and $Z_T$ the
        partition function, i.e., a normalization constant. Now, we can use
        the temperature $T$ as a tuning parameter to adjust the part of the
        distribution we want to gather samples from. Low positive
        temperatures will bias the ``energy'' $S$ towards smaller values because
        decreases in $S$ are always accepted and increases in $S$ are more often
        rejected. For negative $T$ this bias works
        in the opposite way towards larger values of $S$, i.e., larger biconnected
        components in this case.

        For any chosen temperature, the sampling will be restricted to some
        interval determined by the value of $T$. Thus, to obtain the desired
        distribution $P(S)$ over a large range of the support, simulations for
        many different temperatures have to be performed.
        We have to choose the temperatures $T$ in a certain way, to be able to
        reconstruct the wanted distribution $P(S)$ from this data. First,
        we can transform $Q(G)$ into $P(S)$ by summing all realizations
        $G$, which have the same $S$. Hence we obtain with Eq.~\ref{eq:qt}
        \begin{align*}
            P_T(S) &= \sum_{\{G|S(G)=S\}} Q_T(G)\\
                &= \sum_{\{G|S(G)=S\}} \frac{\exp(-S/T)}{Z_T} Q(G)\\
                &= \frac{\exp(-S/T)}{Z_T} P(S).
        \end{align*}
        With this relation we can calculate the wanted, unbiased distribution $P(S)$ from
        measurements of our biased distributions $P_T(S)$. The ratios of all
        constants $Z_T$ can be obtained by enforcing continuity of the
        distribution $P(S)$, i.e.,
        \begin{align*}
            P_{T_j}(S) \ee^{S/T_j} Z_{T_j} = P_{T_i}(S) \ee^{S/T_i} Z_{T_i}.
        \end{align*}
        This requires that our measurements for $P_T(S)$ are at least pairwise
        overlapping such that there is no unsampled region between sampled regions.
        From pairwise overlaps the pairwise ratios $Z_i/Z_j$ can be approximated.
        The absolute value of the $Z_T$ can afterwards be obtained by the
        normalization of $P(S)$.
        Although the size of the largest biconnected
        component $S_2$ is a discrete variable for every finite $N$ and should
        therefore be normalized such that the probabilities of every event
        should sum to one, i.e., $\sum_{i=0}^{N} p(S_2 = i/N) = 1$, we are
        mainly interested in the large $N$ behavior and especially the rate
        function. This limit is continuous and should therefore be treated with
        a different normalization $\int_0^1 p(S_2) \de S_2 = 1$, which we
        approximate for finite $N$ by the trapzoidal rule.
        Anyway, the difference here is just a factor $N$.

        While this technique does usually work quite well and all distributions
        but one exception are obtained with this method, there are sometimes
        first order phase transitions within the finite temperature ensemble,
        rendering it infeasible, or at least very tedious, to acquire values
        inbetween two temperatures. This was a problem here for the modified
        Barab\'asi-Albert graph at the largest simulated graph size $N$. This phenomenon
        is well known and explored in detail in~\cite{Hartmann2011large}.
        We filled this gap by modified Wang-Landau simulations \cite{Wang2001Efficient,Wang2001Determining,Schulz2003Avoiding,Belardinelli2007Fast,Belardinelli2007theoretical}
        with subsequent entropic sampling~\cite{Lee1993Entropic,Dickman2011Complete}.

    \section{Results\label{sec:results}}
        We applied the temperature-based sampling scheme to ER with finite
        connectivities of $c \in \{0.5, 1, 2\}$ and BA with $m = 1.3$
        over practically the whole support $S_2 \in [0,1]$ using around a dozen
        different temperatures for each ensemble and Markov chains of length
        $10^6 N$ to gather enough samples after equilibration and discarding
        correlated samples. Additionally for BA the range $S_2 \in [0.1,0.35]$
        was sampled using Wang-Landau's method and merged into results
        obtained from the temperature based sampling for the remainder of the
        distribution.
        All error estimates for the distributions are
        obtained via bootstrap resampling~\cite{efron1979,young2015} but are
        always smaller than the symbol size and therefore not shown. Error
        estimates for fits are Gnuplot's asymptotic standard errors.


        \begin{figure}[bhtp]
            \centering
            \includegraphics[scale=1]{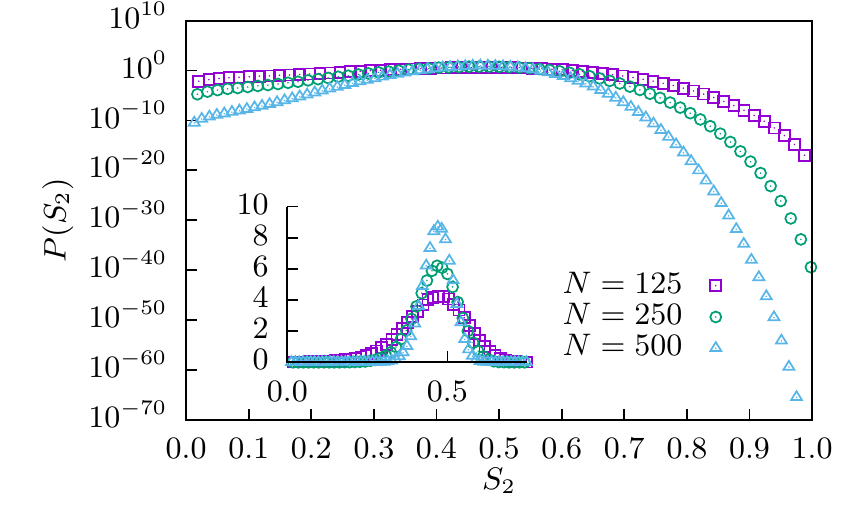}
            \caption{\label{fig:distribution}
                Distributions of the size of the largest biconnected component
                $S_2$ for ER graphs at connectivity $c=2$ and three different
                graph sizes $N$. The main plot shows the distributions in
                logarithmic scale to display the tails, the inset shows the same
                distributions in linear scale, where a concentration around
                the mean value $\avg{S_2}$ (cf.~Fig~\ref{fig:mean}) is visible.
                (For clarity not every bin is visualized.)
            }
        \end{figure}

        Examples for the distribution of the largest biconnected component's
        size for ER graphs at $c=2.0$ are depicted in Fig.~\ref{fig:distribution}
        at three different graph sizes $N$. The inset shows the distribution in
        linear scale, where a concentration with increasing size $N$ around the
        mean value is visible. While the main part of the distribution in the
        inset looks rather symmetric, the tails are obviously not. Also
        it is visible that the tails of the distribution get more and more suppressed
        when increasing the value of $N$.

        \begin{figure}[bhtp]
            \centering
            \includegraphics[scale=1]{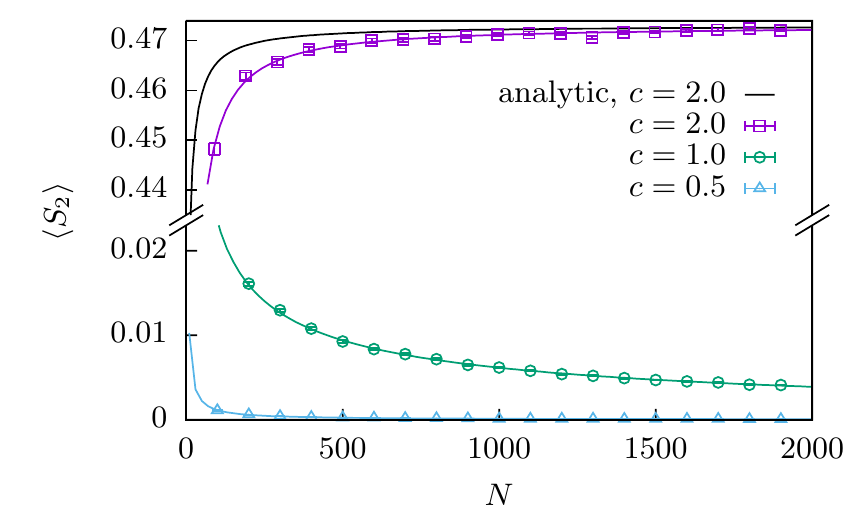}
            \caption{\label{fig:mean}
                Mean size of the largest biconnected component $\avg{S_2}$ for
                different graph sizes $N$. Notice the broken $\avg{S_2}$-axis.
                The black line denotes the analytic expectation for $c=2$ from
                Eq.~\ref{eq:bi_analytic}~\cite{Newman2008Bicomponents}.
                The expectation for $c \le 1$ is $\avg{S_2} = 0$. Fits to a
                power law with offset $\avg{S_2} = a N^{b} + S_2^\infty$ lead to
                $S_2^\infty = -6(8) \cdot 10^{-6}$ for $c=0.5$,
                $S_2^\infty = -0.0013(4)$ for $c=1.0$
                and $S_2^\infty = 0.4729(3)$ for $c=2.0$.
            }
        \end{figure}

        Since the mean size of the biconnected component of ER is
        known for large enough graphs, we will compare the mean sizes of our
        simulations to the analytical expectation.
        Those results are shown in Fig.~\ref{fig:mean}, notice the broken $\avg{S_2}$-axis.
        Apparently at $c=2$ for small sizes $N$ the analytical approximation,
        while close to our measurements, overestimates the size of the biconnected
        component slightly but the relative error diminishes for larger sizes.
        In fact, we extrapolated our measurements to the limit of large $N$ using
        a power-law ansatz $\avg{S_2} = a N^{b} + S_2^\infty$ yielding for
        $c = 0.5$ an offset $S_2^\infty$ compatible within
        errorbars with the expectation $\avg{S_2} = 0$ (exact values in the
        caption of Fig.~\ref{fig:mean}), which is quite remarkable for our ad-hoc
        fit function. The case $c=1$ suggests a negative $S_2^\infty$ close
        to zero, which is probably caused by correction to our assumed scaling
        law. The case $c=2$ seems to converge to the limit of the
        analytical expectation also.
        %


        \begin{figure}[bhtp]
            \centering
            \includegraphics[scale=1]{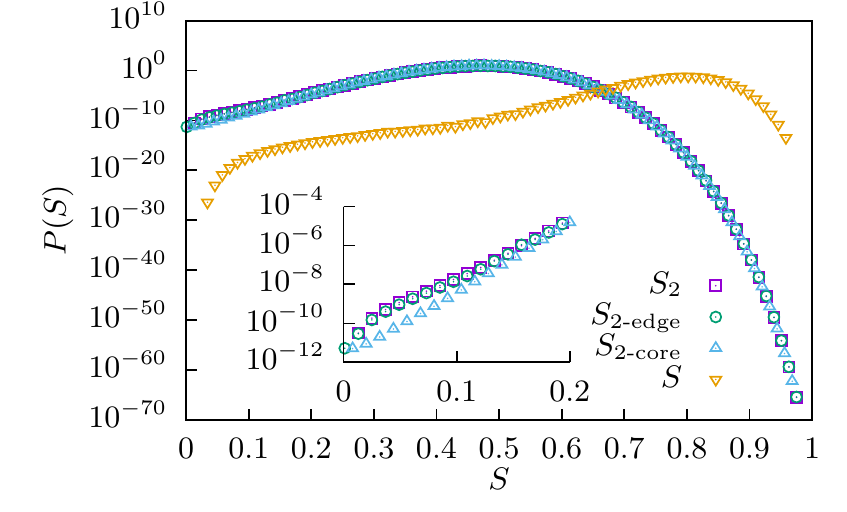}
            \caption{\label{fig:compare}
                Comparison of the relative size of the largest connected
                component $S$ \cite{Hartmann2011large}, the largest $2$-core
                $S_\text{2-core}$ \cite{Hartmann2017Large}, the largest
                bi-edge-connected component $S_\text{2-edge}$ and the
                largest biconnected component $S_2$ for $N=500$ and $c=2$ ER
                graphs. The last three are nearly identical for sizes $S_x \gtrsim 0.2$.
                The inset shows a zoom to the very small components, which is
                the only region, where the three last types deviate considerably
                from one another. For clarity not every data point is visualized.
            }
        \end{figure}

        To compare the sizes of different relevant types of components,
        Fig.~\ref{fig:compare} shows the distributions of
        of the relative size of the largest connected
        component $S$ \cite{Hartmann2011large}, the largest $2$-core
        $S_\text{2-core}$ \cite{Hartmann2017Large}, the largest
        bi-edge-connected component $S_\text{2-edge}$ and the
        largest biconnected component $S_2$ for $N=500$ and $c=2$ ER
        graphs. Interestingly, the distributions $P(S_2)$, $P(S_\text{2-edge})$
        and $P(S_\text{2-core})$ are almost identical and only deviate in
        the region of very small components from each other. As would be expected
        by the order of Eq.~\ref{eq:order}, the probability to find very small
        $2$-cores is lower than to find bi-edge-connected components of the
        same small size, which are again slightly less probable than biconnected
        components of that size. Anyway, when considering ER graphs, which
        exhibit by construction no particular structure, the robustness properties which
        are determined by the biconnected component, can be with very high probability
        readily inferred from the 2-core.

        \begin{figure}[bhtp]
            \centering
            \includegraphics[scale=1]{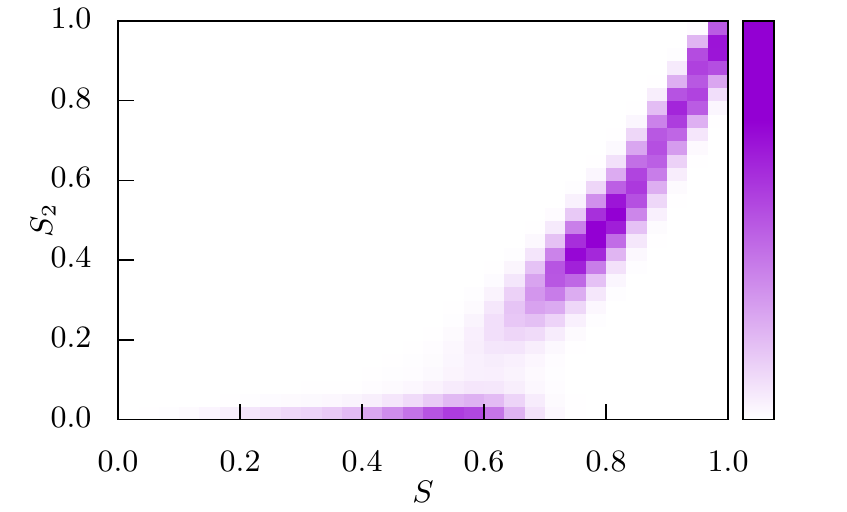}
            \caption{\label{fig:hist}
                Correlation histogram of our raw and biased simulational data.
                A large biconnected component does most probably appear
                in graphs whose connected component is larger than $S \gtrsim 0.6$.
            }
        \end{figure}

        To understand the topology of the instances of very low probabilities
        better, we will look at the correlations of the size of the largest
        connected component $S$ and the largest biconnected component in
        Fig.~\ref{fig:hist}. Note that this histogram does not reflect the
        probabilities, but does count the instances we generated within one of
        our simulations, i.e., data for many different temperatures are shown
        without correction for the introduced bias. Anyway, it is instructive
        to look at this sketch for qualitative understanding. This data is for
        $c=2$ ER at $N=500$. We observe that, even for our biased sampling,
        there are basically no large biconnected components if the connected
        component is smaller than $S \lesssim 0.6$. Above this point, we observe
        larger biconnected components, but generally very few around the size
        $S_2\approx 0.2$.  Above $S \gtrsim 0.6$
        the size of the largest biconnected component is strongly correlated
        with the size of the largest connected component.

        For a qualitative understanding of this behavior, consider the following
        heuristic argument. For the instances without or with very small biconnected
        components, i.e., only short cycles, the graph is basically tree-like.
        Larger biconnected components are then created by connecting two nodes
        of the tree with each other, leading to a cycle which is on average
        roughly half in the order of the size of the tree, leading to the jump in the
        size of biconnected components. The configurations with smaller
        biconnected components are apparently entropically suppressed.

        \begin{figure*}[htb]
            \centering

            \subfigure[\label{fig:rate:c05} $c = 0.5$]{
                \includegraphics[scale=1]{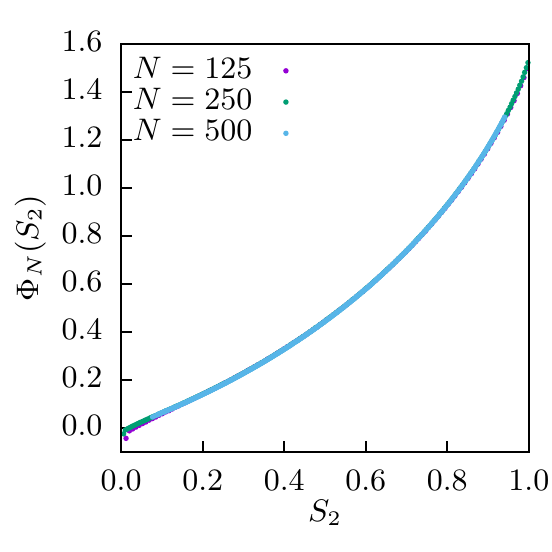}
            }
            \subfigure[\label{fig:rate:c1} $c = 1$]{
                \includegraphics[scale=1]{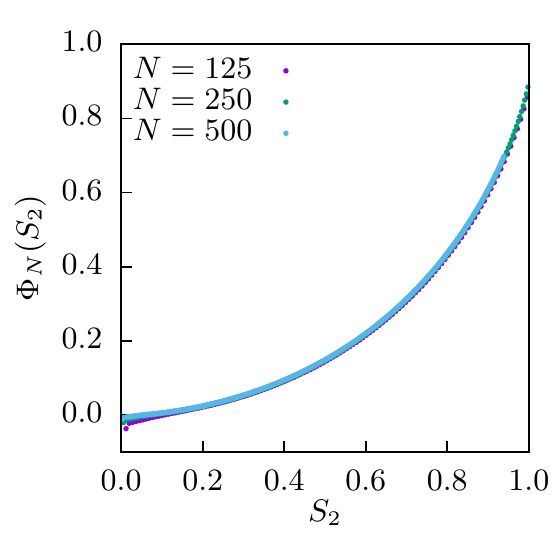}
            }
            \subfigure[\label{fig:rate:c2} $c = 2$]{
                \includegraphics[scale=1]{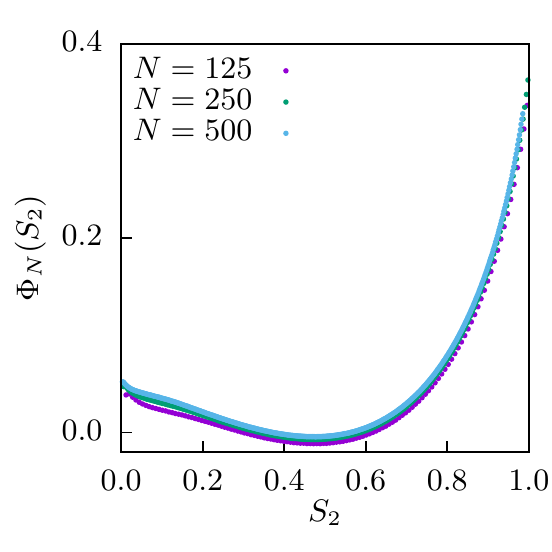}
            }
            \caption{\label{fig:rate}
                Empirical rate function $\Phi_N(S_2)$ for
                multiple graph sizes $N$ and connectivities $c$ of the ER graph
                ensemble.
            }
        \end{figure*}

        Next we will look at the empirical large deviation \emph{rate function}
        of the measured distributions. The rate function $\Phi$ describes the
        behavior of distributions, whose probability density decays
        exponentially in the tails in respect to some parameter $N$. In this
        case, the parameter $N$ is the graph size. For increasing graph size $N$
        the biconnected components which are not typical
        will be exponentially suppressed. To be more precise, the definition
        of the rate function $\Phi(S_2)$ is via
        $P(S_2) = \ee^{-N \Phi(S_2) + o(N)}$ for the large $N$ limit with the
        Landau symbol $o$ for terms of order less than $N$.
        Since we obtained the distributions over most of their support, we can
        access the empirical rate functions $\Phi_N(S_2) = -1/N \log P(S_2)$ for
        finite $N$. Note that the empirical rate functions do contain all
        information of the distribution. If for increasing $N$ they converge
        to the limiting rate
        function, one says the distribution follows the
        \emph{large deviation principle}~\cite{denHollander2000,touchette2009}.

        In Fig.~\ref{fig:rate} the empirical rate functions for ER at different
        connectivities $c$ and for different sizes $N$ are shown.
        Already these comparatively small values of $N$ show remarkably similar
        empiric rate functions and strongly hint at a convergence to a limit
        form. While the rate function in the right tail for larger than typical
        components $S_2$ are already almost indistinguishable, the convergence
        seems a bit slower in the left tail of smaller than typical components.
        This behavior is very similar to the behavior
        of the sizes of the connected component~\cite{Hartmann2011large} and
        the $2$-core~\cite{Hartmann2017Large}. This means that, the large
        deviation principle seems to hold for this distribution.

        \begin{figure}[bht]
            \centering
            \includegraphics[scale=1]{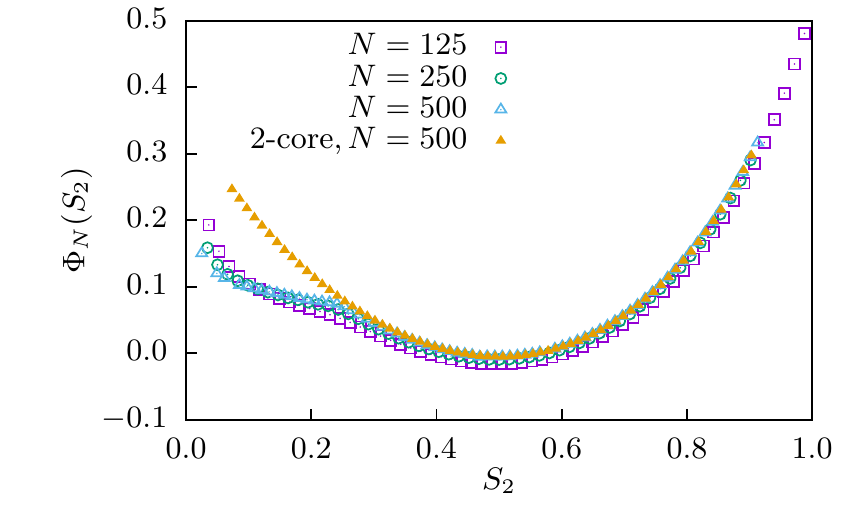}
            \caption{\label{fig:ba_rate}
                Empirical rate function $\Phi_N(S_2)$ for
                multiple graph sizes $N$ of the BA graph ensemble with
                $m = 1.3$.
            }
        \end{figure}

        The rate function of the largest biconnected component of the BA
        ensemble at $m = 1.3$ is shown in Fig.~\ref{fig:ba_rate}. The
        rate function and therefore the distribution does look qualitatively
        similar to the $c = 2$ case of the ER ensemble (cf.~Fig.~\ref{fig:rate:c2}).
        The dip around $S_2 \approx 0.2$ is more pronounced leading to a more
        severe discontinuity in the simulated finite temperature ensemble
        necessitating the use of Wang-Landau sampling. The main qualitative
        difference of the behavior of the two distributions is the behavior for
        very small sizes of the largest biconnected components $S_2$, where the
        empirical rate functions cross each other, hinting at some kind of
        finite size effect suppressing very small biconnected components in
        small graphs.

        In comparison with the rate function of the 2-core, also shown in
        Fig.~\ref{fig:ba_rate}, their difference in the region of very
        small bicomponents respectively 2-cores, which was already observable
        in ER, is very strong in the BA ensemble. Despite those two observables
        are almost indistinguishable in the main region, they show strongly
        different behavior in their overall shape, i.e., the distribution of
        the 2-core seems convex over the region we obtained statistics for.

    \section{Conclusions\label{sec:conclusion}}
        The biconnected component is the fundamental graph-theoretical concept
        which is most related to robustness properties of random networks. Nevertheless,
        the distribution of its size has not been studied before, to our knowledge.
        We used sophisticated sampling methods to obtain the distributions of
        the size of the largest biconnected component $S_2$, for multiple ER
        graph ensembles and
        a modified BA graph ensemble, over a large part of their support.

        For the ER ensemble, looking into the large deviation tails of this
        distribution shows qualitative differences between the size of the
        $2$-core and the biconnected component, which are otherwise not well
        observable. This is even more the case for the BA ensemble where the
        overall shape of the distribution seems to differ. While the 2-core
        distribution seems convex, the distribution of the biconnected component
        shows a ``shoulder''. These qualitative difference, however is only
        apparent below probabilities of $10^{-20}$ and are therefore
        unobservable using conventional methods.

        Further, the empirical rate functions are already for the small sizes
        that we simulated very close to each other
        hinting at a very fast convergence to the limiting form.
        Thus, our results indicate that the large deviation principle
        holds for the numerically obtained distributions.
        This ``well-behaving'' of our numerical results may make
        it promising to address the distribution of the biconnected
        component by analytical means, which has not been done so far to our knowledge.
        Furthermore, it would be interesting to study other network ensembles,
        which are even more relevant for modeling robustness properties, e.g.,
        two-dimensional networks modeling power grids \cite{dewenter2015large} and other
        transportation networks.

    \section{Acknowledgments}
        This work was supported by the German Science Foundation (DFG) through
        the grant HA 3169/8-1.
        We also thank the GWDG (G\"ottingen) for providing computational resources.

    \section{Authors contributions}
        AKH conceived the study, HS wrote the first draft of the manuscript and
        generated most of the new data. All authors contributed ideas,
        simulation data and analysis to this study. All authors were involved
        in the preparation of the manuscript.

    \bibliographystyle{epj}
    \bibliography{lit}

\begin{thebibliography}{48}

\bibitem{newman2003review}
M.E.J. Newman, SIAM Review \textbf{45}, 167 (2003)

\bibitem{dorogovtsev_book2006}
S.N. Dorogovtsev, J.F.F. Mendes, \emph{Evolution of networks: from biological
  nets to the Internet and WWW} (Oxford UNiv.\ Press, 2006)

\bibitem{newman_book2006}
M.~Newman, A.L. Barab\'asi, D.~Watts, \emph{The Structure and Dynamics of
  Networks} (Princeton University Press, 2006)

\bibitem{newman_book2010}
M.~Newman, \emph{Networks: an Introduction} (Oxford University Press, 2010)

\bibitem{barrat_book2012}
A.~Barrat, M.~Barth\'elemy, A.~Vespignani, \emph{Dynamical Processes on Complex
  Networks} (Cambridge University Press, 2012)

\bibitem{Sachtjen2000Disturbances}
M.L. Sachtjen, B.A. Carreras, V.E. Lynch, Phys. Rev. E \textbf{61}, 4877 (2000)

\bibitem{Rohden2012self}
M.~Rohden, A.~Sorge, M.~Timme, D.~Witthaut, Phys. Rev. Lett. \textbf{109},
  064101 (2012)

\bibitem{dewenter2015large}
T.~Dewenter, A.K. Hartmann, New Journal of Physics \textbf{17}, 015005 (2015)

\bibitem{Cohen2000Resilience}
R.~Cohen, K.~Erez, D.~ben Avraham, S.~Havlin, Phys. Rev. Lett. \textbf{85},
  4626 (2000)

\bibitem{lee2006maximum}
D.S. Lee, H.~Rieger, EPL (Europhysics Letters) \textbf{73}, 471 (2006)

\bibitem{ghim2005lethality}
C.M. Ghim, K.I. Goh, B.~Kahng, Journal of Theoretical Biology \textbf{237}, 401
   (2005)

\bibitem{kim2015biconnectivity}
P.~Kim, D.S. Lee, B.~Kahng, Scientific Reports \textbf{5}, 15567 (2015)

\bibitem{erdoes1960}
P.~Erd\H{o}s, A.~R\'enyi, Publ. Math. Inst. Hungar. Acad. Sci. \textbf{5}, 17
  (1960)

\bibitem{watts1998}
D.J. Watts, S.H. Strogatz, Nature \textbf{393}, 440 (1998)

\bibitem{barabasi1999emergence}
A.L. Barab{\'a}si, R.~Albert, Science \textbf{286}, 509 (1999)

\bibitem{albert2000error}
R.~Albert, H.~Jeong, A.L. Barab{\'a}si, Nature \textbf{406}, 378 (2000)

\bibitem{Callaway2000network}
D.S. Callaway, M.E.J. Newman, S.H. Strogatz, D.J. Watts, Phys. Rev. Lett.
  \textbf{85}, 5468 (2000)

\bibitem{Newman2008Bicomponents}
M.E.J. Newman, G.~Ghoshal, Phys. Rev. Lett. \textbf{100}, 138701 (2008)

\bibitem{Norrenbrock2016fragmentation}
C.~Norrenbrock, O.~Melchert, A.K. Hartmann, Phys. Rev. E \textbf{94}, 062125
  (2016)

\bibitem{Kim2013phase}
P.~Kim, D.S. Lee, B.~Kahng, Phys. Rev. E \textbf{87}, 022804 (2013)

\bibitem{biskup2007large}
M.~Biskup, L.~Chayes, S.A. Smith, Random Structures \& Algorithms \textbf{31},
  354 (2007)

\bibitem{Hartmann2011large}
A.K. Hartmann, The European Physical Journal B \textbf{84}, 627 (2011)

\bibitem{Hartmann2017Large}
A.K. Hartmann, The European Physical Journal Special Topics \textbf{226}, 567
  (2017)

\bibitem{denHollander2000}
F.~den Hollander, \emph{Large Deviations} (American Mathematical Society,
  Providence, 2000)

\bibitem{touchette2009}
H.~Touchette, Physics Reports \textbf{478}, 1  (2009)

\bibitem{Hopcroft1973algorithm}
J.~Hopcroft, R.~Tarjan, Commun. ACM \textbf{16}, 372 (1973)

\bibitem{cormen2009introduction}
T.H. Cormen, C.E. Leiserson, R.L. Rivest, C.~Stein, \emph{Introduction to
  algorithms} (MIT press, 2009)

\bibitem{lemon}
B.~Dezs\H{o}, A.~J{\"u}ttner, P.~Kov\'acs, Electronic Notes in Theoretical
  Computer Science \textbf{264}, 23  (2011), proceedings of the Second Workshop
  on Generative Technologies (WGT) 2010

\bibitem{ghoshal2009structural}
G.~Ghoshal, Ph.D. thesis, University of Michigan (2009)

\bibitem{Hartmann2002Sampling}
A.K. Hartmann, Phys. Rev. E \textbf{65}, 056102 (2002)

\bibitem{Wolfsheimer2007local}
S.~Wolfsheimer, B.~Burghardt, A.K. Hartmann, Algorithms for Molecular Biology
  \textbf{2}, 9 (2007)

\bibitem{Fieth2016score}
P.~Fieth, A.K. Hartmann, Phys. Rev. E \textbf{94}, 022127 (2016)

\bibitem{Claussen2015Convex}
G.~Claussen, A.K. Hartmann, S.N. Majumdar, Phys. Rev. E \textbf{91}, 052104
  (2015)

\bibitem{Dewenter2016Convex}
T.~Dewenter, G.~Claussen, A.K. Hartmann, S.N. Majumdar, Phys. Rev. E
  \textbf{94}, 052120 (2016)

\bibitem{schawe2018avoiding}
H.~Schawe, A.K. Hartmann, S.N. Majumdar, Phys. Rev. E \textbf{97}, 062159
  (2018)

\bibitem{schawe2018large}
H.~Schawe, A.K. Hartmann, arXiv preprint arXiv:1808.10698  (2018)

\bibitem{Hartmann2014high}
A.K. Hartmann, Phys. Rev. E \textbf{89}, 052103 (2014)

\bibitem{Engel2004}
A.~Engel, R.~Monasson, A.K. Hartmann, Journal of Statistical Physics
  \textbf{117}, 387 (2004)

\bibitem{hartmann2018distribution}
A.K. Hartmann, M.~M\'ezard, Phys. Rev. E \textbf{97}, 032128 (2018)

\bibitem{Wang2001Efficient}
F.~Wang, D.P. Landau, Phys. Rev. Lett. \textbf{86}, 2050 (2001)

\bibitem{Wang2001Determining}
F.~Wang, D.P. Landau, Phys. Rev. E \textbf{64}, 056101 (2001)

\bibitem{Schulz2003Avoiding}
B.J. Schulz, K.~Binder, M.~M\"uller, D.P. Landau, Phys. Rev. E \textbf{67},
  067102 (2003)

\bibitem{Belardinelli2007Fast}
R.E. Belardinelli, V.D. Pereyra, Phys. Rev. E \textbf{75}, 046701 (2007)

\bibitem{Belardinelli2007theoretical}
R.E. Belardinelli, V.D. Pereyra, The Journal of Chemical Physics \textbf{127},
  184105 (2007)

\bibitem{Lee1993Entropic}
J.~Lee, Phys. Rev. Lett. \textbf{71}, 211 (1993)

\bibitem{Dickman2011Complete}
R.~Dickman, A.G. Cunha-Netto, Phys. Rev. E \textbf{84}, 026701 (2011)

\bibitem{efron1979}
B.~Efron, Ann. Statist. \textbf{7}, 1 (1979)

\bibitem{young2015}
A.P. Young, \emph{Everything You Wanted to Know About Data Analysis and Fitting
  but Were Afraid to Ask}, SpringerBriefs in Physics (Springer International
  Publishing, 2015), ISBN 978-3-319-19050-1

\end{thebibliography}

\end{document}